\documentclass[twocolumn,pre,twoside,showpacs,floatfix]{revtex4}

\usepackage{amsmath}
\usepackage{amssymb}
\usepackage{graphicx}
\usepackage{dcolumn}
\usepackage{bm}
\usepackage{color}
\usepackage{multirow}

\def\epsilon{\varepsilon}
\def\theta{\vartheta}
\def\rho{\varrho}

\begin{document}


\title{Poisson-Boltzmann study of the effective electrostatic interaction between colloids at an electrolyte interface}

\author{Arghya Majee}
\email{majee@is.mpg.de}
\author{Markus Bier}
\email{bier@is.mpg.de}
\author{S.\ Dietrich}
\affiliation{Max-Planck-Institut f\"ur Intelligente Systeme, Heisenbergstr.\ 3, 70569 Stuttgart, Germany}
\affiliation{IV. Institut f\"{u}r Theoretische Physik, Universit\"{a}t Stuttgart, Pfaffenwaldring 57, 70569 Stuttgart, Germany}

\date{\today}

\begin{abstract}
The effective electrostatic interaction between a pair of colloids, both of 
them located close to each other at an electrolyte interface, is studied by employing
the full, nonlinear Poisson-Boltzmann (PB) theory within classical density functional theory. Using a simplified 
yet appropriate model, all contributions to the effective interaction are obtained exactly, 
albeit numerically. The comparison between our results and those obtained within linearized 
PB theory reveals that the latter overestimates these contributions significantly at short inter-particle 
separations. Whereas the surface contributions to the linear and the nonlinear PB results
differ only quantitatively, the line contributions show qualitative differences at short separations. Moreover, 
a dependence of the line contribution on the solvation properties of the two adjacent fluids 
is found, which is absent within the linear theory. Our results are expected to 
enrich the understanding of effective interfacial interactions between colloids.
\end{abstract}


\maketitle


\section{Introduction}

More than a century ago, Ramsden discovered that suspended colloidal particles show a strong affinity for fluid 
interfaces compared to the bulk \cite{Ram03}. This is due to a particle 
induced reduction of the interfacial area between the two fluids. Typically, the resulting decrease in the 
interfacial free energy is much larger than the thermal energy. Thus the attachment of the 
colloids to the interface is almost irreversible and the trapped particles form an effectively two-dimensional system. These 
colloidal monolayers are important for a wide range of industrial and biological proccesses. For example, 
emulsions, including many food emulsions, are stabilized by the adsorption of colloidal particles at liquid-liquid interfaces 
\cite{Dic89, Tam94}; encapsulation and delivery of drugs or nutrients can be achieved through 
colloidosomes \cite{Din02}. Froth flotation, which involves the separation of hydrophilic from hydrophobic particles by attaching 
the latter to air bubbles in a suspension, plays a key role in mineral processing, water purification, oil recovery, bacteria 
separation, and for recycling of plastics \cite{Bin06}. Since the particles are confined only in the vertical direction but are 
free to move along the interfacial plane, the stability of such monolayer structures is, to a large extent, determined by the 
effective inter-particle interaction and therefore a proper understanding of this lateral interaction is 
highly desired.

The effective interaction between particles at an interface is quite different from that present in the bulk. All 
types of interactions, such as electrostatic, magnetic, or van der Waals, which are present in 
the bulk, are also present at the interface, albeit in a different form. On top of that, particles interact via 
deformations of the fluid interface, which can be generated by gravity, electric stress gradients, or magnetic fields \cite{Nik02, 
Oet05, For04, Oet08}. Here, however, we are not concerned with these interface-mediated 
capillary or elastic interactions, but we shall focus only on the electrostatically induced effective
interaction between the particles. The evolution of the studies concerned with the 
electrostatic interaction in this context dates back to 1980, 
when Pieranski reported the two-dimensional crystallization of polystyrene particles at an air-water interface and attributed it 
to a purely repulsive, long-ranged dipolar interaction originating from the asymmetric counterion distribution across the interface 
and acting through air \cite{Pie80}. Hurd confirmed these predictions \cite{Hur85}, based on analytical 
calculations within the framework of linearized PB theory and on the assumption that the 
particles are separated by distances large compared to their radii. Later studies reported a 
weakening of the effective interaction upon increasing the ionic strength of 
the corresponding polar medium which differs from Hurd's prediction \cite{Ave00, Ave02, Hor05, Par08}. As a 
possible explanation this has been linked to the  relatively small amount 
of residual charges present at the particle-oil \cite{Ave00, Ave02, Par08} or the particle-air \cite{Hor05} interface, 
but an unanimous picture is still lacking.

A major simplification used in almost all studies mentioned above as well as in related studies 
consists of considering large inter-particle separations for which both the 
superposition approximation and the linearization of the PB equation are taken to be valid. 
The associated dipolar interaction is also a signature of this key simplification. But for short inter-particle 
distances \cite{Too16}, which is relevant for dense systems or self-organization processes, none of them are 
actually applicable. In a recent 
publication \cite{Maj14}, we have discussed the drawbacks of the superposition assumption not only at short inter-particle 
distances but also at large distances. The assumption concerning the linearization of the PB equation, which 
leads to the well known Debye-H\"uckel (DH) equation, requires the electrostatic energy of a single charge in solution to be much 
smaller than the thermal energy. While this might hold true at distances far from the particle, at short 
distances this is violated for most experimental setups. 
For highly charged colloids, which are also quite common, the situation becomes worse, even at 
relatively large distances. Moreover, nonlinear charge renormalization effects are known to alter the strength of the 
effective interaction potential significantly \cite{Fry07}. Hence, appropriate insight into 
the effective interaction between a pair of particles, especially at small separations, 
remains elusive without including nonlinearity. This formulates the goal of 
the present study.

\begin{figure}[!t]
\centering{\includegraphics[width=8.4cm]{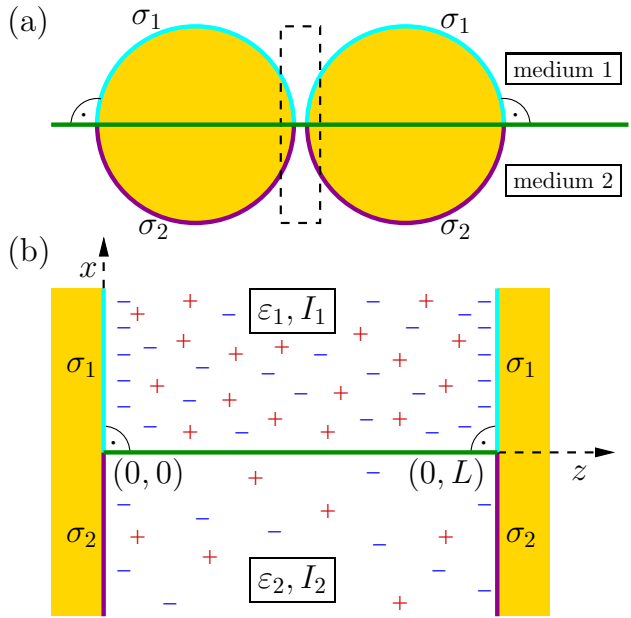}}
\caption{(a) Schematic illustration of two identical spherical particles each of which is partitioned equally 
between the two fluid phases separated by a flat interface (horizontal green line). The particles are close 
to each other with a contact angle of $90^\circ$. (b) Magnified view of the boxed region in (a). Due to the 
small distance between them, the particles are treated as parallel, planar walls 
at a distance $L$. The region between the walls is filled with two immiscible fluids with 
permittivities $\varepsilon_1$, $\varepsilon_2$ and ionic strengths $I_1$, $I_2$ respectively. The surface charge density on 
each wall is $\sigma_1$ ($\sigma_2$) at the contact with medium ``1'' (``2''). The system depicted here 
corresponds to the case in which medium ``1'' is the more polar phase and that 
the walls are positively charged on both sides of the horizontal fluid-fluid interface.}
\end{figure}

Even a mean-field description like the PB equation, which is adequate to describe interfacial 
structures above the atomic scale, poses already a significant challenge. 
Therefore, we numerically solve a simplified model by replacing the spherical particles with planar walls in the 
spirit of the Derjaguin approximation (see Fig.~1) \cite{Rus89}. This simplification is justified 
because we focus on short inter-particle separations only. The validity of such an assumption 
has been verified also for infinitely long cylinders at an oil-water interface \cite{Lyn92}. In 
addition, we consider the particles to float on a flat interface and that they are immersed in both fluid phases 
to the same extent. This corresponds to a  liquid-particle contact angle of $90^\circ$ which is equal or close 
to what one observes in many experimental systems \cite{Ave03, Rei04, Mas10}. In fact, it is known that for high stability of 
particle-stabilized emulsions, the contact angle should not deviate strongly from $90^\circ$ \cite{Ave03}. 
In order to keep the present investigation general and in order to be consistent with 
previous experimental studies \cite{Ave00, Ave02, Par08}, we consider surface charges on both sides of the 
fluid-fluid-interface. The electrostatic problem for this model system is solved by employing the framework of density 
functional theory \cite{Eva79} and the resulting effective interaction energy is divided into two parts: a 
surface contribution, expressed per total surface area, and a line contribution per total length of the two 
three-phase contact lines formed by solid-liquid-liquid coexistence. A comparison of our results 
with those obtained within linear theory reveals both quantitative and qualitative differences concerning the 
effective line interaction energy (Figs.~2(e) and (f)), quantitative differences 
concerning the surface interaction energies (Figs.~2(a)--(d)), and a dependence of the line 
contribution on the solvation properties of the two adjacent fluids which is not captured by 
the linear theory (Fig.~4).


\section{Formalism}

In a three-dimensional Cartesian coordinate system, the walls are considered to be placed at $z=0$ and $z=L$  and the two 
slabs ($x\gtrless0$ with $0\leq z\leq L$) in between are filled with medium ``1'' $(x>0)$ and 
medium ``2'' $(x<0)$, respectively. The walls are chemically identical in nature but the surface charge density 
$\sigma(\mathbf{r})$ on each wall varies depending on the surrounding medium: $\sigma(\mathbf{r})=\sigma_1$ on the upper half 
planes  ($x>0$) and $\sigma(\mathbf{r})=\sigma_2$ on the lower half planes ($x<0$). Since the layering of 
solvent molecules and ions takes place close to the walls and to the interface on the length scale of the bulk correlation length
which, typically, is comparable to the molecular scale and which is much smaller 
than the length scale of interest here, the solvents in both media are taken to be structureless, 
homogeneous, linear dielectric fluids. Therefore, the permittivity $\varepsilon(\mathbf{r})$, which is the 
product of the relative permittivity $\varepsilon_{r}(\mathbf{r})$ and the permittivity $\varepsilon_0$ of vaccum, 
varies steplike at the interface between the fluids: $\varepsilon(\mathbf{r})=\varepsilon_{r,1}\varepsilon_0$ for $x>0$ and 
$\varepsilon(\mathbf{r})=\varepsilon_{r,2}\varepsilon_0$ for $x<0$. The solute is a simple binary salt with 
bulk ionic strength $I(\mathbf{r})=I_1$ ($I_2$) in medium ``1''(``2''). Here, however, we 
consider the nonuniformity of the charge density $e\left(\rho_{+}(\mathbf{r})-\rho_{-}(\mathbf{r})\right)$, with 
$\rho_{\pm}(\mathbf{r})$ denoting the number density of the $\pm$-ions, which varies on 
a length scale of the order of the Debye length which is usually much larger than the molecular length scale. We describe our 
system within the grand canonical ensemble with the ion reservoirs being provided by the bulk phases of both 
media. Considering the ions as point-like objects and ignoring ion-ion correlations, the grand canonical density 
functional corresponding to our system in the units of the thermal energy $k_BT=1/\beta$ is given by
\begin{align}
   \beta\Omega\left[\rho_{\pm}\right] = 
   & \int\limits_Vd^3r\Bigg[\sum\limits_{i=\pm}\rho_i\left(\mathbf{r}\right)
     \Bigg\{\ln\left(\frac{\rho_i\left(\mathbf{r}\right)}{\zeta_i}\right)-1\notag\\
   & +\beta V_i\left(\mathbf{r}\right)\Bigg\}+\frac{\beta\mathbf{D}
     \left(\mathbf{r},\left[\rho_\pm\right]\right)^2}{2\varepsilon\left(\mathbf{r}\right)}\Bigg],
\label{eq:1}
\end{align}
where $\zeta_{\pm}$ are the fugacities of the two species of ions, $\mathbf{D}$ is the electric displacement 
field, and the integration volume $V$ is the space enclosed by the two walls. The first line of Eq.~(\ref{eq:1}) represents the 
entropic ideal gas contribution of the ions. The first term in the second line describes the ion-solvent interaction expressed by 
an external potential $V_{\pm}(\mathbf{r})$ acting on the ions \cite{Bie12}. The last term corresponds to the ion-ion Coulomb 
interaction. First, one determines the equilibrium density profiles $\rho_{\pm}^{\text{eq}}$, which minimize the 
grand potential in Eq.~(\ref{eq:1}). Second, these equilibrium profiles are inserted into the 
grand potential functional in order to infer the equilibrium grand potential $\beta\Omega^{\text{eq}}(L)=\beta
\Omega\left[\rho_{\pm}^{\text{eq}}\right]$. In the course of this minimization process, one 
encounters the relation
\begin{align}
   \rho_{\pm}^{\text{eq}}(\mathbf{r}) =
   I(\mathbf{r})\exp\left\{\mp\beta e\left(\Phi\left(\mathbf{r},[\rho_{\pm}^{\text{eq}}]\right)-\Phi_b(\mathbf{r})\right)\right\}
   \label{eq:1.5}
\end{align}
with the electrostatic potential $\Phi(\mathbf{r},[\rho_{\pm}^{\text{eq}}])$ satisfying the nonlinear Poisson-Boltzmann
equation
\begin{align}
   \nabla^2\left(\beta e\Phi\left(\mathbf{r}\right)\right)
   &=\kappa(\mathbf{r})^2\sinh\left\{\beta e\left(\Phi\left(\mathbf{r}\right)-\Phi_b(\mathbf{r})\right)\right\}
 \label{eq:2}
\end{align}
everywhere except for $x=0$. We note that from here onwards for reasons of brevity the functional 
dependence of $\Phi$ on $\rho_{\pm}^{\text{eq}}$ is not indicated explicitly. The associated boundary conditions 
are: (i) the electrostatic potential should remain finite for $x\rightarrow\pm\infty$, (ii) both the electrostatic potential and 
the normal component of the electric displacement field should be continuous at the fluid interface, i.e., 
$\Phi(x=0^+)=\Phi(x=0^-)$ and $\varepsilon_{r,1}\partial_x\Phi(x=0^+) = \varepsilon_{r,2}\partial_x\Phi(x=0^-)$, and (iii) in order to 
satisfy global charge neutrality, the normal component of the electric displacement field should match the surface charge 
densities at both walls, i.e., $\varepsilon(\mathbf{r})\partial_z\Phi(z=0)=\sigma(\mathbf{r})$ and 
$\varepsilon(\mathbf{r})\partial_z\Phi(z=L)=-\sigma(\mathbf{r})$. In Eq.~(\ref{eq:2}), 
$\kappa(\mathbf{r})=\sqrt{8\pi\ell_BI(\mathbf{r})/\varepsilon_r(\mathbf{r})}$ is the inverse Debye screening length with the 
Bjerrum length $\ell_B=e^2/(4\pi\varepsilon_0k_BT)$, $e>0$ is the elementary charge, and $\Phi_b(\mathbf{r})$ represents the 
electrostatic potential in the bulk of the two media which is defined such that $\Phi_b(\mathbf{r})=0$ in medium ``1'' 
($x>0$) and $\Phi_b(\mathbf{r})=\Phi_D$ in medium ``2'' ($x<0$). $\Phi_D$ originates from a difference in the 
solubilities of the ions in the two fluids and is called Donnan potential or Galvani potential difference between the two media 
\cite{Bag06}.

In order to numerically determine $\Phi(\mathbf{r})$, which solves Eq.~(\ref{eq:2}) and fulfills 
the boundary conditions (i)--(iii), we use a Rayleigh-Ritz-like finite element method (fem) based on the 
minimization of the functional
\begin{align}
   \beta\Omega_{\text{fem}}\left[\Phi\right] = 
   &\int\limits_{V}d^3r\Bigg[2I(\mathbf{r})\cosh\left\{\beta e \left(\Phi(\mathbf{r})-\Phi_b(\mathbf{r})\right)\right\}\notag\\
   &+\frac{\beta\varepsilon(\mathbf{r})}{2}\left\{\left(\partial_x\Phi(\mathbf{r})\right)^2+\left(\partial_z\Phi(\mathbf{r})
    \right)^2\right\}\Bigg]\notag\\
   &-\int\limits_{\partial V}d^2r\sigma(\mathbf{r})\Phi(\mathbf{r}),
\label{eq:3}
\end{align}
with $\partial V$ indicating the boundaries of the integration volume $V$. It can be shown that the minimum 
$\beta\Omega_{\text{fem}}^{\text{min}}(L)$ of Eq.~(\ref{eq:3}) is related to the equilibrium grand potential 
$\beta\Omega^{\text{eq}}(L)$ by $\beta\Omega^{\text{eq}}(L)=
-\beta\Omega_{\text{fem}}^{\text{min}}(L)$. In order to study the effect of nonlinearity, we expand the function 
$\cosh\left\{\beta e \left(\Phi(\mathbf{r})-\Phi_b(\mathbf{r})\right)\right\}$ in Eq.~(\ref{eq:3}) in a 
Taylor series:
\begin{align}
   \beta\Omega_{\text{fem}}^{(n)}\left[\Phi\right] = 
   &\int\limits_{V}d^3r\Bigg[2I(\mathbf{r})\sum\limits_{k=0}^n\frac{\left\{\beta e \left(\Phi(\mathbf{r})-\Phi_b(\mathbf{r})
    \right)\right\}^{2k}}{(2k)!}\notag\\
   &+\frac{\beta\varepsilon(\mathbf{r})}{2}\left\{\left(\partial_x\Phi(\mathbf{r})\right)^2+\left(\partial_z\Phi(\mathbf{r})
    \right)^2\right\}\Bigg]\notag\\
   &-\int\limits_{\partial V}d^2r\sigma(\mathbf{r})\Phi(\mathbf{r}),
\label{eq:4}
\end{align}
where $n$ describes the degree of the nonlinearity. For $n=1$, it reduces to the linearized PB problem and for 
$n\rightarrow\infty$ it corresponds to the full nonlinear problem (see Eq.~(\ref{eq:2})). First, we find the 
equilibrium profiles for the electrostatic potential $\Phi^{\text{eq}}$ which minimize the functional in 
Eq.~(\ref{eq:4}) and then we insert it back into Eq.~(\ref{eq:4}) in order to calculate the 
grand potential $\beta\Omega^{(n)\text{eq}}(L)=-\beta\Omega_{\text{fem}}^{(n)\text{min}}(L)$. The latter 
includes nine distinct contributions:
\begin{align}
   \Omega^{(n)\text{eq}}(L) = 
   &\sum\limits_{i\in\{1,2\}}\left[\Omega_{b,i}V_i+\left(\gamma_i+\omega_{\gamma,i}(L)\right)A_i\right]\notag\\
   &+\gamma_{1,2}A_{1,2}+\left(\tau+\omega_\tau(L)\right)\ell,
\label{eq:5}
\end{align}
where $\Omega_{b,i}$ is the bulk grand potential density (i.e., the negative osmotic pressure of the ions) in medium 
$i\in\{1,2\}$, $\gamma_i$ is the surface tension of a single wall in contact with medium $i$, $\omega_{\gamma,i}(L)$ is
the surface interaction energy per total area of the two walls at distance $L$ in contact with medium $i$, 
$\gamma_{1,2}$ is the interfacial tension between the two fluid media, $\tau$ is the line tension of a single three-phase 
(solid-liquid-liquid) contact line, $\omega_{\tau}(L)$ is the line interaction energy per total line length of 
two three-phase contact lines at a distance $L$, $V_i$ is the volume of the slab between the two walls filled 
with medium $i$, $A_i$ is the total surface area of the two walls in contact with medium $i$, and $\ell$ is the total length of 
the two three-phase contact lines. In order to separate all these parts, we solve the following additional problems: (i) a single 
medium in the absence of any wall; the corresponding grand potential density is obtained by setting 
$\Phi(\mathbf{r})=\Phi_b(\mathbf{r})$ and $\sigma(\mathbf{r})=0$ in Eq.~(\ref{eq:4}) which leads to 
$\Omega_{b,i}=-2I_i/\beta$, (ii) two fluid media forming an 
interface in the absence of any walls ($\sigma(\mathbf{r})=0$ in Eq.~(\ref{eq:4})), (iii) one homogeneously charged wall in 
contact with a single semi-infinite fluid medium, (iv) two homogeneously charged walls in 
contact with a single fluid medium in between, and (v) a single charged wall in contact with two immiscible 
semi-infinite fluids forming an interface along with a single three-phase contact line. Finally, a systematic subtraction of one 
interaction potential from another, which corresponds to one of the above mentioned problems, allows one to 
extract all individual contributions in Eq.~(\ref{eq:5}). We note that, upon 
construction, all $L$-dependent contributions, i.e., $\omega_{\gamma,i}(L)$ and $\omega_\tau(L)$, vanish 
individually in the limit $L\rightarrow\infty$.

\begin{figure*}[!t]
\centering{\includegraphics[width=13.28cm]{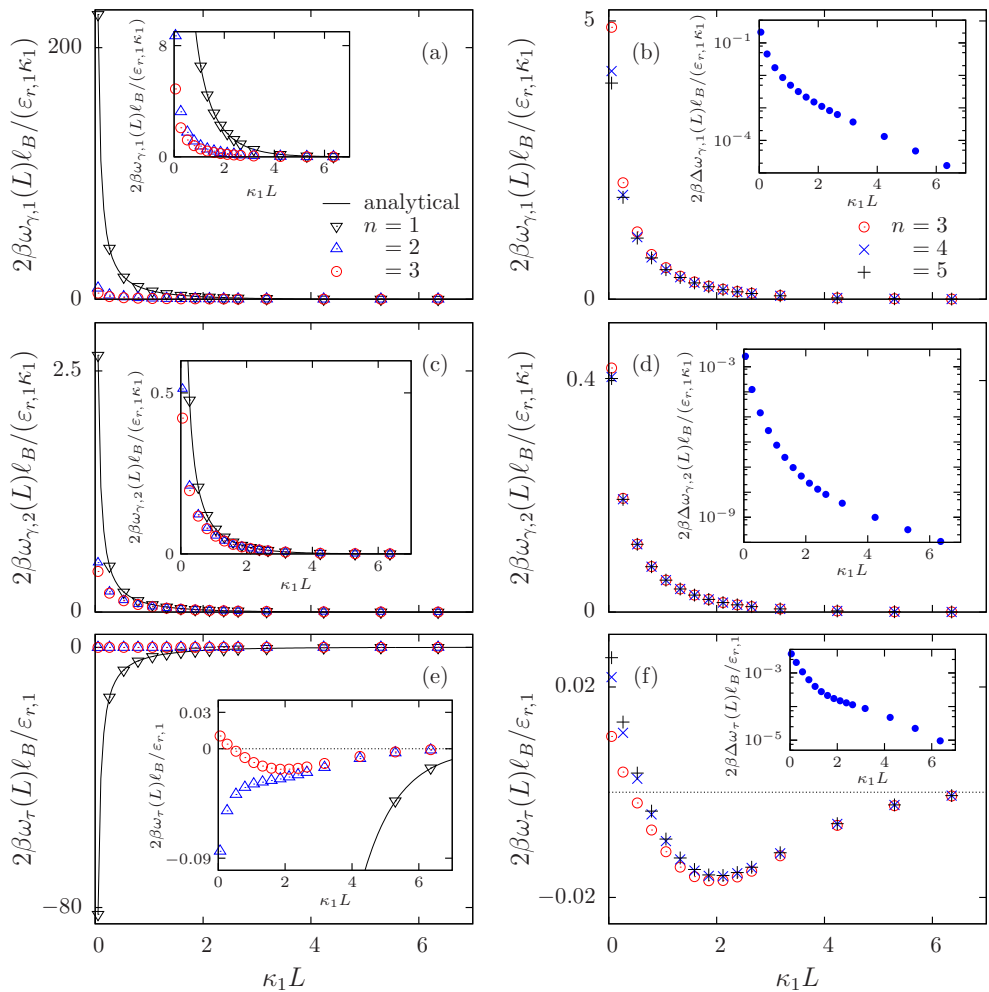}}
\caption{(a)--(d) Reduced surface interaction energies $\omega_{\gamma,i}(L)$ per total area of 
the two walls in contact with medium $i\in\{1,2\}$ and (e)--(f) reduced line interaction energy 
$\omega_{\tau}(L)$ per total length of the three-phase contact lines as functions of the wall separation $L$ 
(in units of the Debye screening length $1/\kappa_1$) for various degrees $n$ 
of nonlinearity (see Eq.~(5)). For these plots the standard set of parameters ($\sigma=0.1$, $I=0.85$, 
$\varepsilon=62/72$, $b_1\approx0.23$, $\beta e\Phi_D=1$, see the main text) corresponding to a water+lutidine 
mixture with KBr at $T=313\,\mathrm{K}$ is used. In the left panels, values for $1\leq n\leq 3$ are compared
whereas in the right panels the values for $3\leq n\leq 5$ are compared. For $n=1$, our numerical results match perfectly with the 
corresponding analytical solutions (black solid line) taken from Ref.~\cite{Maj14}. With increasing 
$n$, the magnitude of both $\omega_{\gamma,1}(L)$ and $\omega_{\gamma,2}(L)$ drop significantly at short separations, but 
the interactions remain repulsive everywhere. However, $\omega_{\tau}(L)$ shows qualitative changes upon 
increasing $n$. Whereas $\omega_{\tau}(L)$ is attractive at all separations both in the linear case ($n=1$) 
and for $n=2$, it is repulsive at short distances for $n\geq3$. Due to the huge difference of magnitudes between the results 
obtained from the linear and the nonlinear theory, in the left panels magnified views are provided as insets. 
Since the differences $\Delta\omega_{\gamma,i}(L)=\omega_{\gamma,i}^{n=5}(L)-\omega_{\gamma,i}^{n=4}(L)$ with $i\in\{1,2\}$ and 
$\Delta\omega_{\tau}(L)=\omega_{\tau}^{n=5}(L)-\omega_{\tau}^{n=4}(L)$ are very small, the absolute values of these differences 
are displayed as insets in the right panels.}
\end{figure*}

\begin{figure}[!t]
\centering{\includegraphics[width=7.5cm]{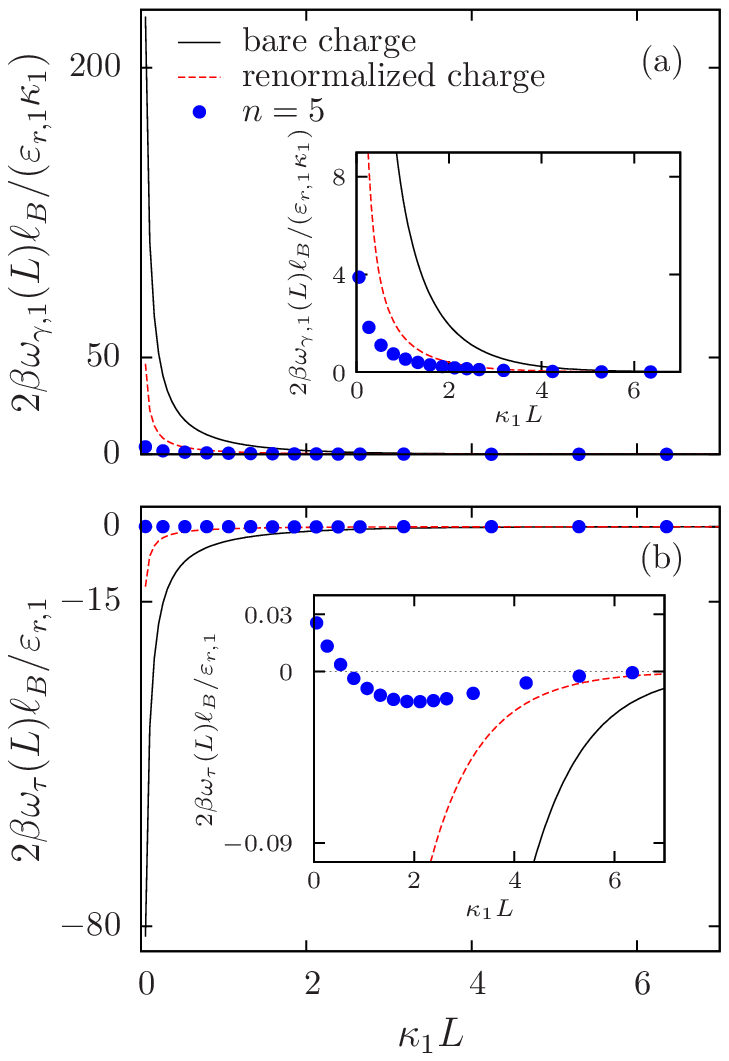}}
\caption{(a) Reduced surface interaction energy $\omega_{\gamma,1}(L)$ per total area of 
the two walls in contact with medium ``1'' and (b) reduced line interaction energy $\omega_{\tau}(L)$ per total 
length of the three-phase contact lines as functions of the scaled wall separation $\kappa_1L$ for three
different cases: (i) The black solid lines correspond to the analytical solutions taken from Ref.~\cite{Maj14}
within linearized PB theory and with bare surface charge densities at the walls. (ii) The red dashed lines also correspond
to the analytical solutions taken from Ref.~\cite{Maj14} within linearized PB theory but with the renormalized surface
charge density in medium ``1''. (iii) The blue filled circles correspond to the numerical solutions within 
the nonlinear PB theory with the degree $n=5$ of nonlinearity. As in Fig.~2, the standard set of parameters is used 
for the plots (see the main text). The quantitative differences between linear and nonlinear results 
decrease upon using the renormalized surface charges. But still a significant mismatch remains to be 
present and the qualitative features obtained within the nonlinear theory remain unexplained by the linear theory
even after taking into account the charge renormalization effect.}
\end{figure}


\section{Results and Discussion}

\subsection{$L$-dependent interactions}

\begin{figure}[!t]
\centering{\includegraphics[width=7.0cm]{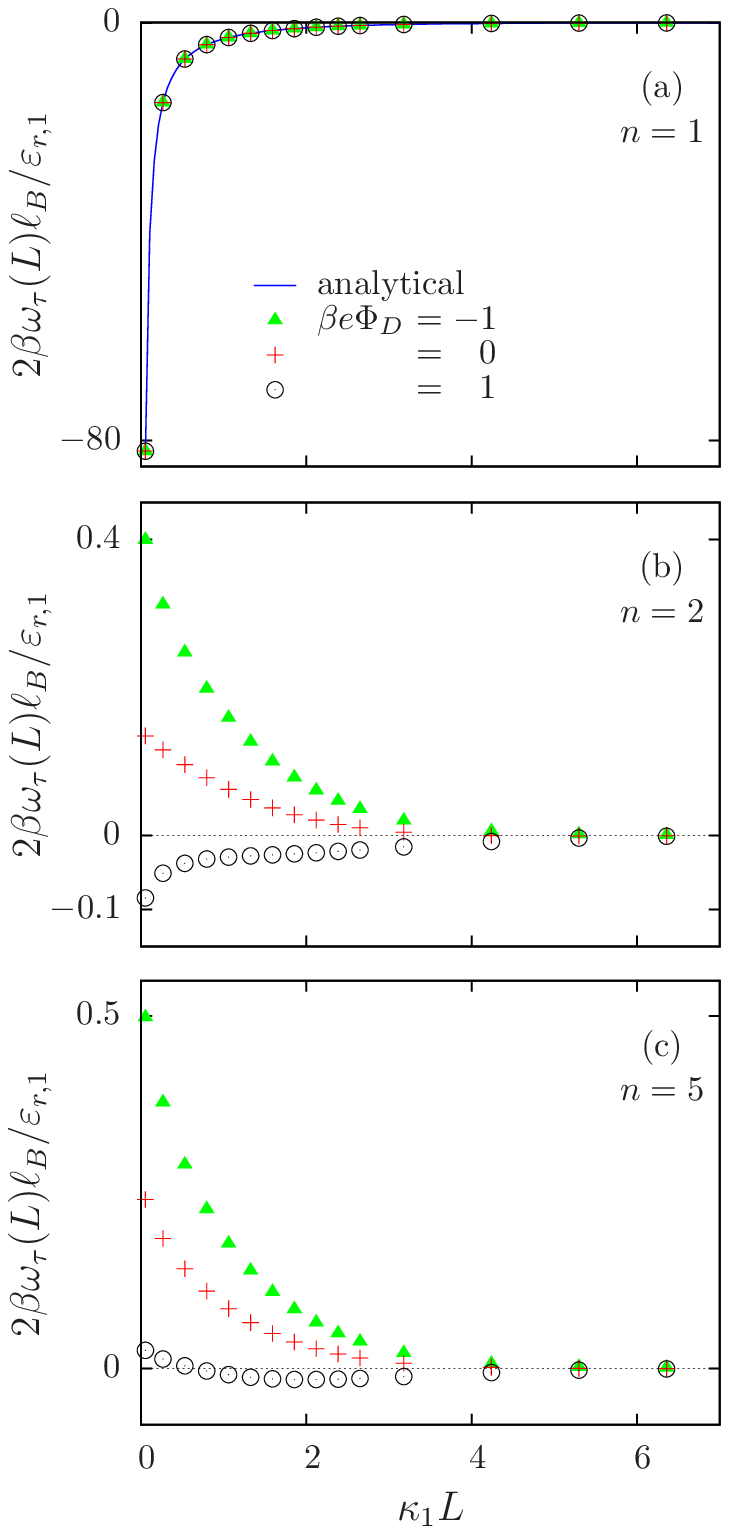}}
\caption{Line interaction energy $\omega_{\tau}(L)$ per total length of the three-phase contact lines expressed in units of 
$\varepsilon_{r,1}/(2\beta\ell_B)$ as a function of the distance $L$ (in units of the Debye
length $1/\kappa_1$) for three values of the dimensionless Donnan potential $\beta e\Phi_D$. (a) Within the 
linear theory ($n=1$), $\omega_{\tau}(L)$ is insensitive to varying $\beta e\Phi_D$ which is in accordance 
with the analytical results obtained within the linearized PB theory \cite{Maj14}. 
(b) However, for the degree $n=2$ of the nonlinearity, $\omega_{\tau}(L)$ 
varies upon changing $\beta e\Phi_D$. For $n\geq4$ the line interaction curves $\omega_{\tau}(L)$ basically do 
not depend on $n$ anymore. (c) For $n=5$, $\omega_{\tau}(L)$ exhibits a minimum for $\beta e\Phi_D=1$ (which 
corresponds to its value within the standard set of parameters) but it is purely repulsive for $\beta e\Phi_D=-1$ 
and $0$ (even for $n=2$) whereas the linear theory predicts an attractive interaction.}
\end{figure}

In the following, we discuss all $L$-dependent contributions to the effective interaction between the walls as a function of 
the scaled wall separation $\kappa_1L$. Accordingly, both $\omega_{\gamma,i}(L)$ and $\omega_{\tau}(L)$ are 
rendered dimensionless by expressing them in units of $\varepsilon_{r,1}
\kappa_1/(2\beta\ell_B)$ and $\varepsilon_{r,1}/(2\beta\ell_B)$, respectively. If done so, the dimensionless 
free parameters for our system turn out to be $\sigma=\sigma_2/\sigma_1$, $I=I_2/I_1$, $\varepsilon=
\varepsilon_{r,2}/\varepsilon_{r,1}$, $\beta e\Phi_D$, and $b_1=\kappa_1\ell_{GC}^{(1)}$, 
where $\ell_{GC}^{(1)}=e\varepsilon_{r,1}/
(2\pi\lvert\sigma_1\rvert\ell_B)$ is the Gouy-Chapmann length for medium ``1''. First, we discuss the results for a standard 
set of parameters ($\sigma_{\text{std}}=0.1$, $I_{\text{std}}=0.85$, $\varepsilon_{\text{std}}=62/72$, $b_1\approx0.23$, 
$\beta e\Phi_D=1$) which corresponds to a typical experimental setup, e.g., a water-lutidine (2,6-dimethylpyridine) mixture with 
NaI salt ($1\,\mathrm{mM}$ in the aqueous phase) at temperature $T=313\,\mathrm{K}$ in contact with polystyrene 
walls exhibiting, in the aqueous phase, a surface charge density of $0.1\,e/\mathrm{nm^2}$ 
\cite{Ave02, Bie12, Gra93, Ine94, Ram58, Lid98, Gal92}. The resulting interaction energies are presented 
in Fig.~2. In each of the three plots, the symbols correspond to various values of the degree of nonlinearity 
$n$ in Eq.~(\ref{eq:4}) and the solid lines correpond to the analytical solutions taken from Ref.~\cite{Maj14}. 
Figures~2(a) and (b) show the reduced surface interaction energy $\omega_{\gamma,1}(L)$ between the 
walls in contact with medium ``1'' for a varying distance $L$ between the walls. While the interaction remains 
repulsive in all cases, obviously the linear theory overestimates the strength of the 
interaction at short distances. As expected, for $n=1$ (which corresponds to the linear theory), our numerical results match with 
the corresponding analytical results over the whole range of separations considered here. The 
most significant changes take place upon increasing the degree of nonlinearity $n$ from $n=1$ (linear theory) to $n=2$,
whereas for $n\geq4$ basically no changes occur upon increasing $n$ further. The surface 
interaction $\omega_{\gamma,1}(L)$ within the linear theory differs by orders of magnitude from the one within the nonlinear 
theory. For example, $\omega_{\gamma,1}(\kappa_1L=0.05)$ for $n=1$ is larger compared to that obtained for $n=5$ by almost a 
factor of $60$. This discrepancy diminishes gradually with increasing separation distance, but even for 
$\kappa_1L=2$, a factor of almost $10$ is still present. Similar features are obtained for $\omega_{\gamma,2}(L)$ 
as well (Figs.~2(c) and (d)). However, in this case the mismatch is less severe
because medium ``2'' is the less polar phase, for which the electrostatic interaction is 
expected to be weaker compared to the one for the more polar phase. Figures~2(e) and (f) 
compare the line interaction energies $\omega_\tau(L)$ corresponding to various degrees of 
nonlinearity $n$. For them the linear theory predicts a monotonically weakening,
attractive interaction upon increasing separations between the walls. Upon 
increasing the degree of nonlinearity $n$, the strength of this interaction decreases and $\omega_\tau(L)$ 
becomes nonmonotonic for $n\geq3$ forming a minimum at a distance $\kappa_1L\approx2$ which, 
for typical Debye lengths of $10\,\mathrm{nm}$ corresponding to a $1\,\mathrm{mM}$
($\approx0.0006\,\mathrm{nm^{-3}}$) aqueous solution, is well above the molecular scale ($<1\,\mathrm{nm}$). 
Regardless of the type of interaction discussed above, for $\kappa_1L\leq8.5$ its magnitude 
within the linear theory is at least one order of magnitude larger than within the nonlinear theory. 

\begin{figure*}[!t]
\centering{\includegraphics[width=12.5cm]{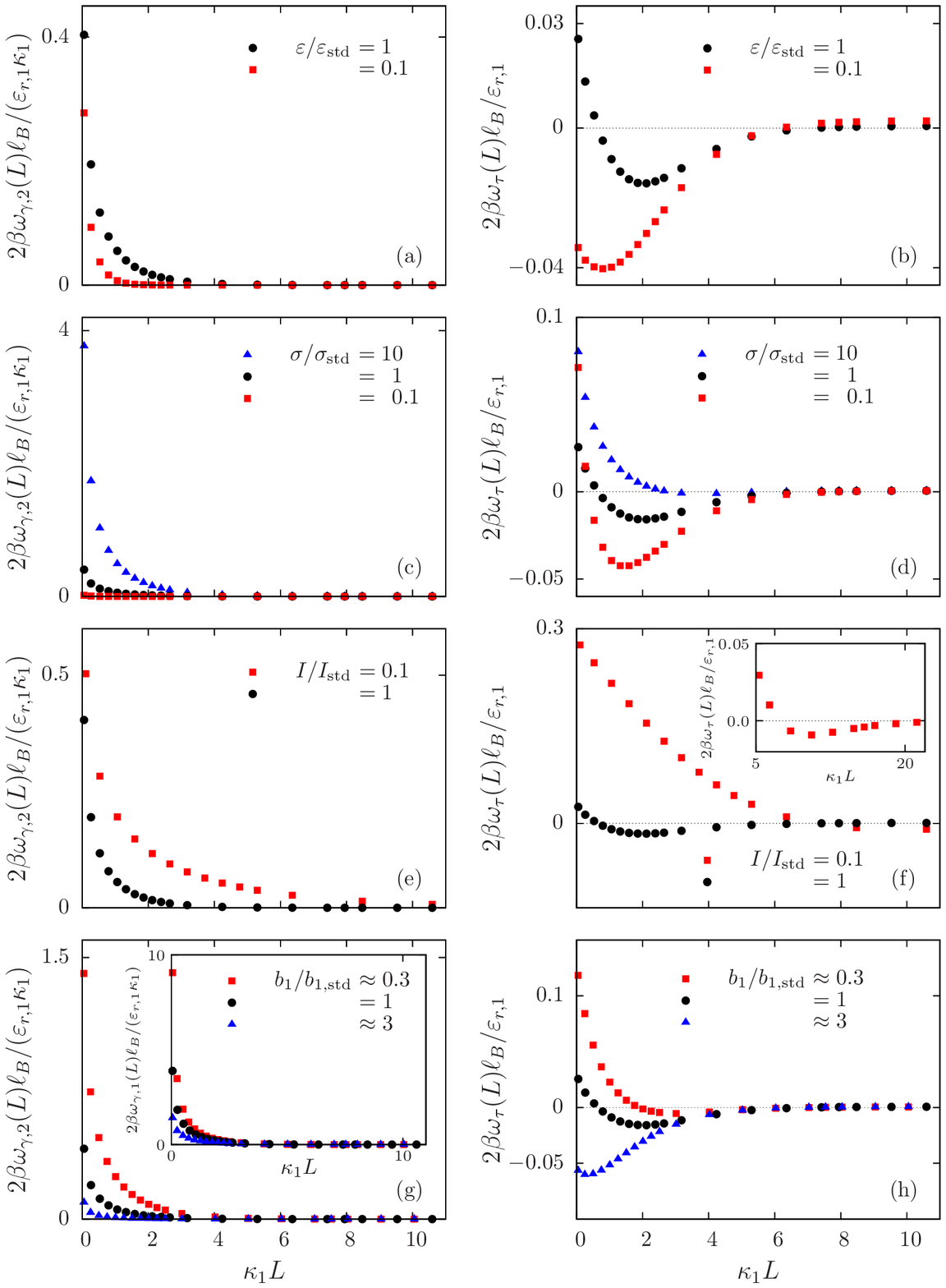}}
\caption{Left panels: Surface interaction energy $\omega_{\gamma,2}(L)$ per total area of the two walls in contact with 
medium ``2'' expressed in units of $\varepsilon_{r,1}\kappa_1/(2\beta\ell_B)$ as a function of the 
rescaled wall separation $\kappa_1L$ for various sets of the free dimensionless parameters 
$\varepsilon$, $\sigma$, $I$, and $b_1$, respectively. Although $\omega_{\gamma,2}(L)$ is always repulsive, an increase in 
$\varepsilon$ or $\sigma$ strengthens the repulsion whereas an increase in $I$ or $b_1$ weakens 
it. Inset of panel (g): Scaled surface interaction energy per total area of the two walls in contact with medium ``1'' as a 
function of $\kappa_1L$. As one can infer from the plots, $\omega_{\gamma,1}$ and $\omega_{\gamma,2}$ behave 
similarly upon varying $b_1$. Right panels: Line interaction energy $\omega_{\tau}(L)$ per total length of the 
three-phase contact lines in units of $\varepsilon_{r,1}/(2\beta\ell_B)$ as a function 
of $\kappa_1L$ for various sets of the dimensionless parameters. Compared with the standard set 
of parameters, the observed minimum of $\omega_{\tau}(L)$ becomes deeper and 
shifts towards smaller separations upon decreasing $\varepsilon$ or $\sigma$. The opposite 
trend occurs upon decreasing $I$ or $b_1$. The inset of panel (f) shows the variation of $\omega_{\tau}(L)$ for a larger range 
of $\kappa_1L$.}
\end{figure*}

Within the linearized PB theory it is a common practice to use renormalized instead of bare
surface charge densities in order to capture the correct asymptotic behavior of the electrostatic potential 
at \textit{large} distances \cite{Boc02}.
However, in the present study we are interested in the opposite limit of \textit{short} distances. Still, it is
interesting to see the effects of charge renormalization on $\omega_{\gamma,i}(L)$ and $\omega_{\tau}(L)$. 
This implies a replacement of the bare surface charge density ($\sigma_i$) with an effective charge
density ($\pm\sigma_{\text{eff}}^{(i)}$) if $\sigma_{\text{eff}}^{(i)}<\lvert\sigma_i\rvert$. The analytic expression
for the effective surface charge density is known for a single charged wall in contact with a semi-infinite
electrolyte solution and is given by $\sigma_{\text{eff}}^{(i)}=e\kappa_i\varepsilon_{r,i}/(\pi\ell_B)$, 
$i\in\{1,2\}$, with $\kappa(x>0)=\kappa_1$ and $\kappa(x<0)=\kappa_2$ \cite{Boc02}. For the above mentioned 
standard set of parameters, we have calculated separately $\sigma_{\text{eff}}^{(i)}$ for a single wall in contact with
medium $i\in\{1,2\}$; it turns out that $\sigma_{\text{eff}}^{(1)}<\sigma_1$ whereas $\sigma_{\text{eff}}^{(2)}>\sigma_2$.
Accordingly, we have only replaced $\sigma_1$ by $\sigma_{\text{eff}}^{(1)}$ keeping $\sigma_2$ the same. The corresponding
results are presented in Fig.~3. As expected, both $\omega_{\gamma,1}(L)$ and $\omega_{\tau}(L)$ decrease in
magnitude compared to the case of bare surface charge densities, but a significant quantitative mismatch
compared with the full nonlinear behavior is still present. However, the results corresponding to the linear
theory with bare or renormalized charge densities show the same qualitative behavior. Therefore, features
like the minimum in Fig.~3(b) cannot be explained by simply using surface charge renormalization within
the linear PB theory.

The strong reduction in strength of both the surface and the line parts raises the question 
concerning the relevance of the electrostatic interaction. In order to answer this, we compare 
our results with the van der Waals (vdW) interaction which is calculated in terms of the 
Hamaker constant \cite{Ham37}. For two flat surfaces made of polystyrene and interacting accross pure water, the Hamaker constant 
$A$ is reported to lie in the interval $[0.95, 1.3]\times10^{-20}\,\mathrm{J}$ \cite{Isr85}. Even for the maximal 
value of $A$, the vdW attraction energy $-A/(12\pi L^2)$ per cross-sectional area due to the 
two surfaces is either comparable (for $L<2.5\,\mathrm{nm}$) or less by at least one order of magnitude compared to the 
corresponding electrostatic part. It is important to note that salting the water further 
decreases the vdW contribution slightly due to the electrostatic screening effect \cite{Isr85}. 
Therefore, the electrostatic part still contributes significantly to the total effective interaction. We are not 
aware of any such data regarding the line contribution to the interaction.

In the following we discuss the effects of varying the free parameters of our system. 
To this end, one of the five dimensionless parameters $\beta e\Phi_D$,
$\varepsilon$, $\sigma$, $I$, or $b_1$ is varied at a time, keeping the remaining ones fixed. First, we consider 
the dimensionless Donnan potential $\beta e\Phi_D$. According to the linear theory \cite{Maj14}, all three $L$-
dependent interaction energies, i.e., $\omega_{\gamma,1}(L)$, $\omega_{\gamma,2}(L)$, and $\omega_{\tau}(L)$, are 
independent of $\Phi_D$, which is confirmed by our numerical results. Moreover, this holds for the surface interactions 
$\omega_{\gamma,1}(L)$ and $\omega_{\gamma,2}(L)$ for arbitrary degrees $n$ of nonlinearity. In contrast, 
the line part $\omega_\tau(L)$ exhibits a completely different behavior (see Fig.~4). For $n=1$, $\omega_\tau(L)$ is attractive 
at all separations $L$ and independent of the value of $\Phi_D$. However, for $n\geq2$ the line interaction 
$\omega_\tau(L)$ does depend on $\Phi_D$ (Fig.~4(b)). Starting from $n=4$, $\omega_\tau(L)$ hardly changes upon 
increasing the order of nonlinearity $n$. Figure~4(c) displays the corresponding results for $n=5$. For 
$\beta e\Phi_D\in\{-1,0\}$, $\omega_\tau(L)$ differs significantly in magnitude but it is repulsive everywhere, 
whereas for $\beta e\Phi_D=1$, which corresponds to its value in the standard set of parameters, it is repulsive 
at close separations $L$, but attractive further away. It is also worth noting that for $\beta e\Phi_D\in\{-1,0\}$, the 
predictions of the linear theory are qualitatively wrong at all separations $L$.

Effects of changing the remaining parameters $\varepsilon$, $\sigma$, $I$, and $b_1$ are shown in Fig.~5. Although 
here we present the results corresponding to the degree of nonlinearity $n=5$, for each set of parameters, the 
interaction energy curves basically do not change for $n\geq4$. Therefore, in view of 
potential future studies, we conclude that it is sufficient to truncate the Taylor series in Eq.~(\ref{eq:5}) at 
$n=4$. We change $\varepsilon$, $\sigma$, and $I$ by changing $\varepsilon_2$, $\sigma_2$, and $I_2$, 
respectively, while keeping their counterparts for medium ``1'' fixed. Consequently, $\omega_{\gamma,1}(L)$ does not change 
upon these variations; it changes only upon varying $b_1$. As shown by 
Fig.~5, although $\omega_{\gamma,2}(L)$ remains always repulsive (left column of panels), both its magnitude 
and its range depend on the above parameters. An increase in $\varepsilon$ or $\sigma$ 
strengthens the repulsion whereas an increase in $I$ or $b_1$ weakens the repulsion. The 
range of this repulsive energy is set by the Debye length (or equivalently the ionic strength) in medium ``2"; 
for lower ionic strength the range increases. Upon changing $b_1$, $\omega_{\gamma,1}(L)$ behaves 
similarly as $\omega_{\gamma,2}(L)$ (see inset in Fig.~5(g)). Regarding the line interaction 
energy $\omega_\tau(L)$, the observed minimum for the standard set of parameters becomes deeper and shifts towards smaller 
separations upon decreasing $\varepsilon$ or $\sigma$. On the other hand, it becomes more shallow and shifts 
to a larger separation distance $L$ upon decreasing $I$ or $b_1$. Since in our system medium ``2'' 
is the less polar phase and since the standard values for $\varepsilon$ and $I$ are already close to unity, we 
do not have the option to increase these two quantities further. 
Nonetheless, the concomitant trends can be easily inferred from the data presented here.

What remains to be discussed is the relative importance of the line interaction $\omega_{\tau}$ as
compared to the surface interactions $\omega_{\gamma,1}$ and $\omega_{\gamma,2}$. According to Eq.~(\ref{eq:5}), the 
surface contributions scale with the area of the walls whereas the line contribution scales with the length of the 
three-phase contact lines. Hence, for sufficiently large surface area, the surface 
contributions eventually dominate. However, depending on 
the system parameters, the line contribution in Eq.~(\ref{eq:5}) can dominate over the surface contributions even
for typical sizes of the colloid particles. 
For example, we consider a generic system which consists of a water-lutidine (2,6-dimethylpyridine) 
mixture with NaI salt at temperature $T=313\,\mathrm{K}$, occupying the space between the two charged walls with an effective area  
$A_i=25\times10^4\,\mathrm{nm^2}$ ($i\in\{1,2\}$). The effective length of the three-phase contact line is taken as 
$\ell=1000\,\mathrm{nm}$. The ionic concentration in the more polar water-rich phase is given by $I_1=0.1\,\mathrm{mM}$
and that for the less polar lutidine-rich phase is given by $I_2=0.085\,\mathrm{mM}$. The relative permittivities
of the water- and lutidine-rich phases are given by $\varepsilon_{r,1}=72$ and $\varepsilon_{r,2}=62$, respectively.
The walls carry a surface charge density of $\sigma_1=0.01\,e/\mathrm{nm^2}$ in contact with the water-rich phase and 
a surface charge density of $\sigma_2=0.0001\,e/\mathrm{nm^2}$ in contact with the lutidine-rich phase. Therefore the 
system is characterized by the following dimensionless parameters: $I=0.85$, $\sigma=0.01$, $\varepsilon=62/72$, 
$b_1=0.72$, and $\beta e\Phi_D=1$. For such a system, the line contribution for a wall separation $L\approx90\,\mathrm{nm}$ 
(or $\kappa_1L\approx3$), corresponding to the position of the minimum of $\omega_{\tau}$, is $\omega_{\tau}\ell\approx-27\,k_BT$, whereas 
the surface contributions are $\omega_{\gamma,1}A_1\approx5\,k_BT$ and $\omega_{\gamma,2}A_2\approx0.02\,k_BT$. 
Systems with other media are characterized by a different set of values for $\varepsilon$, $I$, and $\beta e\Phi_D$. For an 
oil with a smaller value of $\varepsilon_{r,2}$ the line interaction is expected to be more prominent because, with 
decreasing $\varepsilon$, $\omega_{\gamma,2}$ decreases whereas the minimum in $\omega_{\tau}$ becomes deeper (see Figs.~5(a) 
and (b)). The ratio $I$ can change in at least two ways: An increase of $I_1$ leads to a faster decay of $\omega_{\gamma,1}$
whereas a decrease of $I_2$ shifts the minimum of $\omega_\tau$ to larger values of $\kappa_1L$ (see Fig.~5(f)). Therefore, 
in both cases, the relative importance of the line interaction $\omega_{\tau}$ with respect to the surface interactions 
increases upon increasing $I$. An increase in the value of $\beta e\Phi_D$ will also deepen the minimum in $\omega_\tau$
leading to a stronger line interaction (see Fig.~4). It is important to note that $\omega_{\gamma,1}$ and $\omega_{\gamma,2}$
remain unaffected due to a change in the Donnan potential $\beta e\Phi_D$.

\subsection{$L$-independent interactions}

\begin{table*}
\centering
\resizebox{0.85\textwidth}{!}{
\begin{tabular}{|c|ll|l|l|l|l|l|}
\hline
& \multicolumn{2}{c|}{\multirow{2}{*}{analytical}} & \multicolumn{5}{c|}{numerical}\\
\cline{4-8}
&&&\multicolumn{1}{c|}{$n=1$} & \multicolumn{1}{c|}{$n=2$} & \multicolumn{1}{c|}{$n=3$} & \multicolumn{1}{c|}{$n=4$} & \multicolumn{1}{c|}{$n=5$} \\ \hline
\multirow{4}{*}{standard} & $\gamma_{s,1}$ & $=0.4398\,\ k_\text{B}T/\mathrm{nm^2}$ & $0.4398$ & $0.2992$ & $0.2812$ & $0.2781$ & $0.2777$ \\
 & $\gamma_{s,2}$ & $=0.01514\,\ k_\text{B}T/\mathrm{nm^2}$ & $0.01514$ & $0.01504$ & $0.01504$ & $0.01504$ & $0.01504$ \\
 & $\gamma_{1,2}$ & $=-0.002621\,\ k_\text{B}T/\mathrm{nm^2}$ & $-0.002621$ & $-0.002635$ & $-0.002635$ & $-0.002635$ & $-0.002635$ \\
 & $\tau$ & $=-0.5669\,\ k_\text{B}T/\mathrm{nm}$ & $-0.5686$ & $-0.06757$ & $-0.03333$ & $-0.02752$ & $-0.02657$ \\ \hline
\multirow{4}{*}{$\beta e\Phi_D=0$} & $\gamma_{s,1}$ & $=0.4398\,\ k_\text{B}T/\mathrm{nm^2}$ & $0.4398$ & $0.2992$ & $0.2812$ & $0.2781$ & $0.2777$ \\
 & $\gamma_{s,2}$ & $=0.005141\,\ k_\text{B}T/\mathrm{nm^2}$ & $0.005140$ & $0.005037$ & $0.005035$ & $0.005035$ & $0.005035$ \\
 & $\gamma_{1,2}$ & $=0\,\ k_\text{B}T/\mathrm{nm^2}$ & $0$ & $0$ & $0$ & $0$ & $0$ \\
 & $\tau$ & $=-0.9509\,\ k_\text{B}T/\mathrm{nm}$ & $-0.9520$ & $-0.2655$ & $-0.2067$ & $-0.1963$ & $-0.1947$ \\ \hline
\multirow{4}{*}{$\beta e\Phi_D=-1$} & $\gamma_{s,1}$ & $=0.4398\,\ k_\text{B}T/\mathrm{nm^2}$ & $0.4398$ & $0.2992$ & $0.2812$ & $0.2781$ & $0.2777$ \\
 & $\gamma_{s,2}$ & $=-0.004859\,\ k_\text{B}T/\mathrm{nm^2}$ & $-0.004860$ & $-0.004963$ & $-0.004965$ & $-0.004965$ & $-0.004965$ \\
 & $\gamma_{1,2}$ & $=-0.002621\,\ k_\text{B}T/\mathrm{nm^2}$ & $-0.002621$ & $-0.002635$ & $-0.002635$ & $-0.002635$ & $-0.002635$ \\
 & $\tau$ & $=-1.335\,\ k_\text{B}T/\mathrm{nm}$ & $-1.3366$ & $-0.5040$ & $-0.4250$ & $-0.4108$ & $-0.4084$ \\ \hline
\multirow{4}{*}{$\varepsilon/\varepsilon_{\text{std}}=0.1$} & $\gamma_{s,1}$ & $=0.4398\,\ k_\text{B}T/\mathrm{nm^2}$ & $0.4398$ & $0.2992$ & $0.2812$ & $0.2781$ & $0.2777$ \\
 & $\gamma_{s,2}$ & $=0.0263\,\ k_\text{B}T/\mathrm{nm^2}$ & $0.02624$ & $0.02423$ & $0.02406$ & $0.02405$ & $0.02404$ \\
 & $\gamma_{1,2}$ & $=-0.001210\,\ k_\text{B}T/\mathrm{nm^2}$ & $-0.001211$ & $-0.001224$ & $-0.001224$ & $-0.001224$ & $-0.001224$ \\
 & $\tau$ & $=-0.1401\,\ k_\text{B}T/\mathrm{nm}$ & $-0.1424$ & $0.009198$ & $0.01742$ & $0.01876$ & $0.01899$ \\ \hline
\multirow{4}{*}{$\sigma/\sigma_{\text{std}}=10$} & $\gamma_{s,1}$ & $=0.4398\,\ k_\text{B}T/\mathrm{nm^2}$ & $0.4398$ & $0.2992$ & $0.2812$ & $0.2781$ & $0.2777$ \\
 & $\gamma_{s,2}$ & $=0.6141\,\ k_\text{B}T/\mathrm{nm^2}$ & $0.6140$ & $0.4317$ & $0.4081$ & $0.4037$ & $0.4030$ \\
 & $\gamma_{1,2}$ & $=-0.002621\,\ k_\text{B}T/\mathrm{nm^2}$ & $-0.002621$ & $-0.002635$ & $-0.002635$ & $-0.002635$ & $-0.002635$ \\
 & $\tau$ & $=-0.1127\,\ k_\text{B}T/\mathrm{nm}$ & $-0.1133$ & $-0.07228$ & $-0.07077$ & $-0.07052$ & $-0.07046$ \\ \hline
\multirow{4}{*}{$\sigma/\sigma_{\text{std}}=0.1$} & $\gamma_{s,1}$ & $=0.4398\,\ k_\text{B}T/\mathrm{nm^2}$ & $0.4398$ & $0.2992$ & $0.2812$ & $0.2781$ & $0.2777$ \\
 & $\gamma_{s,2}$ & $=0.001051\,\ k_\text{B}T/\mathrm{nm^2}$ & $0.001051$ & $0.001051$ & $0.001051$ & $0.001051$ & $0.001051$ \\
 & $\gamma_{1,2}$ & $=-0.002621\,\ k_\text{B}T/\mathrm{nm^2}$ & $-0.002621$ & $-0.002635$ & $-0.002635$ & $-0.002635$ & $-0.002635$ \\
 & $\tau$ & $=-0.7615\,\ k_\text{B}T/\mathrm{nm}$ & $-0.7634$ & $-0.1469$ & $-0.09980$ & $-0.09171$ & $-0.09039$ \\ \hline
\multirow{4}{*}{$I/I_{\text{std}}=0.1$} & $\gamma_{s,1}$ & $=0.4398\,\ k_\text{B}T/\mathrm{nm^2}$ & $0.4396$ & $0.2988$ & $0.2808$ & $0.2777$ & $0.2772$ \\
 & $\gamma_{s,2}$ & $=0.02626\,\ k_\text{B}T/\mathrm{nm^2}$ & $0.02626$ & $0.02425$ & $0.02409$ & $0.02407$ & $0.02407$ \\
 & $\gamma_{1,2}$ & $=-0.001210\,\ k_\text{B}T/\mathrm{nm^2}$ & $-0.001210$ & $-0.001223$ & $-0.001223$ & $-0.001223$ & $-0.001223$ \\
 & $\tau$ & $=-0.4251\,\ k_\text{B}T/\mathrm{nm}$ & $-0.4283$ & $-0.2093$ & $-0.1973$ & $-0.1957$ & $-0.1954$ \\ \hline
\multirow{4}{*}{$b_1/b_{1,\text{std}}\approx3$} & $\gamma_{s,1}$ & $=0.1391\,\ k_\text{B}T/\mathrm{nm^2}$ & $0.1391$ & $0.1251$ & $0.1241$ & $0.1240$ & $0.1240$ \\
 & $\gamma_{s,2}$ & $=0.01163\,\ k_\text{B}T/\mathrm{nm^2}$ & $0.01163$ & $0.01162$ & $0.01162$ & $0.01162$ & $0.01162$ \\
 & $\gamma_{1,2}$ & $=-0.008287\,\ k_\text{B}T/\mathrm{nm^2}$ & $-0.008288$ & $-0.008332$ & $-0.008332$ & $-0.008332$ & $-0.008332$ \\
 & $\tau$ & $=0.02634\,\ k_\text{B}T/\mathrm{nm}$ & $0.02573$ & $0.03243$ & $0.03284$ & $0.03287$ & $0.03287$ \\ \hline
\multirow{4}{*}{$b_1/b_{1,\text{std}}\approx0.3$} & $\gamma_{s,1}$ & $=1.391\,\ k_\text{B}T/\mathrm{nm^2}$ & $1.391$ & $0.6061$ & $0.5098$ & $0.4860$ & $0.4798$ \\
 & $\gamma_{s,2}$ & $=0.02626\,\ k_\text{B}T/\mathrm{nm^2}$ & $0.02626$ & $0.02425$ & $0.02408$ & $0.02407$ & $0.02407$ \\
 & $\gamma_{1,2}$ & $=-0.0008287\,\ k_\text{B}T/\mathrm{nm^2}$ & $-0.0008288$ & $-0.0008332$ & $-0.0008332$ & $-0.0008332$ & $-0.0008332$ \\
 & $\tau$ & $=-8.295\,\ k_\text{B}T/\mathrm{nm}$ & $-8.305$ & $-0.4430$ & $-0.1653$ & $-0.1129$ & $-0.09931$ \\ \hline
\end{tabular}
}
\caption{Values of the surface tensions $\gamma_{s,1}$ and $\gamma_{s,2}$ at the walls in contact with medium 
``1'' and ``2'', respectively, interfacial tension $\gamma_{1,2}$, and line tension $\tau$ at the three-phase contact 
lines for various sets of parameters $\varepsilon$, $\sigma$, $I$, and $b_1$. The second column provides the 
values for the linearized theory calculated by using Eqs.~(\ref{eq:5.5})--(\ref{eq:6}). 
The remaining columns provide the values corresponding to various orders 
of the nonlinearity $n$ considered in Eq.~(\ref{eq:4}). As expected, the values for $n=1$ agree well 
with the analytical results. For higher orders $n$ of the nonlinearity, the magnitudes of the 
surface tensions $\gamma_{s,i}$ decrease noticeably, whereas the absolute value of the interfacial tension $\gamma_{1,2}$ 
increases slightly. However, the quantity, which is most sensitive with respect to $n$, is the line tension 
$\tau$. In some cases it exhibits variations of several orders of magnitude as function of $n$; 
upon increasing the degree of nonlinearity $n$ it can increase, decrease, or even change sign. 
For $n\geq4$ the dependences on $n$ level off.}
\end{table*}

In this subsection we discuss the remaining, $L$-independent contributions $\gamma_{s,i}$, $\gamma_{1,2}$, 
and $\tau$ in Eq.~(\ref{eq:5}). Analytical expressions for these quantities can be obtained within linear
theory (i.e., degree of nonlinearity $n=1$), following the same procedure as the one presented in Ref.~\cite{Maj14}:
\begin{align}
  \gamma_{s,i}&=\frac{\sigma_i^2}{2\kappa_i\varepsilon_i}+\sigma_i\Phi_b,\qquad i\in\{1,2\},
  \label{eq:5.5}\\
  \gamma_{1,2}&=-\frac{\kappa_1\kappa_2\varepsilon_1\varepsilon_2\Phi_D^2}{\kappa_1\varepsilon_1+\kappa_2\varepsilon_2},
  \label{eq:5.6}
\end{align}
and
\begin{widetext}
 \begin{align}
  \tau=\!\left(\frac{\Phi_D}{1+\kappa\varepsilon}\right)\!\!\left(\frac{\kappa\varepsilon\sigma_1}{\kappa_1}-\frac{\sigma_2}
       {\kappa_2}\right)
       \!\!-\!\frac{\sigma_1^2}{\kappa_1^2\varepsilon_1}\!\int\limits_0^\infty \!dx\!\bigg[\frac{\frac{\sigma}{\varepsilon}
       \frac{1}{x^2\pi^2+\kappa^2}\!-\!\frac{1}{x^2\pi^2+1}}{1\!+\!\frac{\sqrt{x^2\pi^2+1}}{\varepsilon\sqrt{x^2\pi^2+\kappa^2}}}
       \frac{1}{\sqrt{x^2\pi^2+1}}
       +\frac{\frac{\sigma}{x^2\pi^2+1}\!-\!\frac{\sigma^2}{\varepsilon}\frac{1}{x^2\pi^2+\kappa^2}}
       {1\!+\!\frac{\varepsilon\sqrt{x^2\pi^2+\kappa^2}}{\sqrt{x^2\pi^2+1}}}\frac{1}{\sqrt{x^2\pi^2+\kappa^2}}\bigg],
 \label{eq:6}
 \end{align}
\end{widetext}
where $\varepsilon_i=\varepsilon_{r,i}\varepsilon_0$, $i\in\{1,2\}$ and $\kappa=\kappa_2/\kappa_1$. In Table I the values 
resulting from these expressions along with the ones obtained numerically are listed for each set 
of parameters. As one can see, the numerical values for $n=1$ (3rd column) agree well with the analytical results (2nd column). 
Whereas the values vary as functions of the degree of nonlinarity $n$, they do not change significantly for $n\geq4$. For 
weaker interactions (e.g., in the less polar phase) this convergence is even 
more rapid. In line with the $L$-dependent interactions $\omega_{s,i}(L)$, the magnitudes of the surface 
tensions $\gamma_{s,i}$ decrease upon increasing the degree of nonlinearity $n$. On the other hand, the absolute value of the 
interfacial tension $\gamma_{1,2}$ increases slightly with increasing order of nonlinearity $n$ and, as expected, it is 
independent of the surface charge densities at the walls. For an ionic strength of $1\,\mathrm{mM}$ in the 
aqueous phase, the predicted values for the interfacial tension also agrees well with those
obtained in earlier studies \cite{Gue39, Bie08}. In contrast, the line tension $\tau$ is
most sensitive to the degree of nonlinearity. Depending on the values chosen for the free 
parameters, $\tau$ can either increase or decrease with increasing $n$; in some cases (e.g., for
$b_1/b_{1,\text{std}}\approx0.3$) it varies by two orders of magnitude upon 
increasing $n$. For $\varepsilon/\varepsilon_{\text{std}}=0.1$, it even changes sign due to the presence of the
nonlinearity. The values for our expressions for the line tension $\tau$ are either of the  order of 
$1\,\mathrm{pN}$ or slightly less, which is consistent with values reported in the literature \cite{Bre98, Get98, 
Bau99, Pom00, Mug02}. By considering variations of the parameter $b_1$ it can be inferred from Tab.~I that the line tension 
$\tau$ increases upon increasing the ionic strength $I$. This is not in contradiction to the decrease of $\tau$ upon increasing 
$I$ close to a wetting transition, as reported in Ref.~\cite{Iba16}: In fact, in that study only contact angles 
of $<50^\circ$ occurred, and the observed decrease of $\tau$ upon increasing the ionic strength $I$ has been 
found to become smaller for increasing contact angles (Fig.~5(b) in Ref.~\cite{Iba16}); 
in contrast, in the present study there are no wetting transitions and the contact angle is large, i.e., $90^\circ$.


\section{Conclusion} 

Within the framework of nonlinear Poisson-Boltzmann theory, we have addressed the issue regarding the 
electrostatic interaction between a pair of identical, charged walls at distance $L$, 
separated by two immiscible electrolyte solutions forming a flat interface (Fig.~1). Our numerical findings 
demonstrate that for small $L$ the linear theory overestimates all $L$-dependent contributions 
to the total electrostatic interaction by at least one order of magnitude (Fig.~2). Within the nonlinear theory 
the qualitative trends of the effective surface and line interaction potentials as functions of all system 
parameters have been discussed (Figs.~4 and 5). Whereas the variations as functions of the degree 
$n$ of nonlinearity of the surface and interfacial contributions (i.e., the surface interaction 
energy, the surface tension, and the interfacial tension) are only of a 
quantitative character, the line contributions (line interaction energy and line tension) show 
a qualitatively different behavior within the nonlinear theory as compared to 
the corresponding predictions of the linear theory (Figs.~2, 4, 5 and Tab.~I). For example, while the linear theory predicts a 
monotonically decreasing, attractive interaction between the two three-phase contact lines, nonlinearity changes 
it to a repulsive one at close separations which turns attractive at large distances $L$ 
(Figs.~2(e) and (f)). These differences between the linear and the nonlinear theory cannot be explained by 
using, within the linear theory, a simple charge renormalization procedure (Fig.~3). 
Depending on the parameters, the line tension is found to change sign (Tab.~I). Moreover, a 
dependence of the line interaction energy on the solvation properties of the two media (described in terms of the Donnan 
potential) is present only within the nonlinear theory (Fig.~4). The degree of nonlinearity is 
given by truncating the series expansion of the right-hand side of the Poisson-Boltzmann 
equation~(\ref{eq:2}). The ensuing results indicate that it is sufficient to consider only the first few 
terms in this series. The system we have studied is expected to mimic the situation of two colloidal particles 
being trapped very close to each other at a liquid interface. Depending on the interaction 
among each other, particles are known to form stable, unstable, or even mesostructures at a liquid 
interface \cite{Bin06}. In the unstable situation, particles aggregate to form fractal structures. Whereas for a stable monolayer 
to form, long-ranged interactions between the particles are required. On the
other hand, the formation  of fractal structures requires the total interaction potential to be short-ranged 
and characterized by a minimum at very short distances. Accordingly, our findings are 
particularly relevant with respect to this aspect. Thus, we expect our results 
for a pair of colloids to contribute towards a better understanding of the formation and 
stability of many-body colloidal monolayers trapped at fluid interfaces.


\end{document}